\documentclass[a4paper,11pt]{article}
\usepackage{pos}
\title{Simulation studies for the Mini-EUSO detector}
\ShortTitle{Simuation studies for the MiniEUSO detector}
\author{H.~Miyamoto$^{1,2*}$, F.~Fenu$^{1,2}$, D.~Barghini$^{1,2,12}$, M.~Battisti$^{1,2}$, A.~Belov$^3$, M.~Bertaina$^{1,2}$, 
F.~Bisconti$^{1,2}$, 
R.~Bonino$^{1,2}$, 
G.~Cambie$^{6,7}$, 
F.~Capel$^{8}$, 
M.~Casolino$^{6,7,11}$, 
I.~Churilo$^{13}$, 
T.~Ebisuzaki$^{11}$, 
C.~Fuglesang$^{8}$, 
A.~Golzio$^{1,2}$, P.~Gorodetzky$^{4}$, 
F.~Kajino$^{18}$, P.~Klimov$^{3}$,
M.~Manfrin$^{1,2}$,
L.~Marcelli$^{6,7}$, 
W.~Marsza\l$^{14}$, 
M.~Mignone$^{1}$, 
E.~Parizot$^{4}$, 
P.~Picozza$^{6,7}$,
L.W.~Piotrowski$^{11}$, Z.~Plebaniak$^{14}$, G.~Pr\'{e}v\^{o}t$^{4}$, 
E.~Reali$^{7}$, M.~Ricci$^{17}$
N.~Sakaki$^{11}$, K.~Shinozaki$^{14}$, G.~Suino$^{1,2}$, J.~Szabelski$^{14}$, Y.~Takizawa$^{11}$\\
}
\affiliation{
$^1$INFN Turin, Italy, 
$^2$University of Turin, Department of Physics, Italy, 
$^3$SINP, Lomonosov Moscow State University, Moscow, Russia., 
$^4$APC, Univ Paris Diderot, CNRS/IN2P3, CEA/Irfu, Obs de Paris, Sorbonne Paris Cit\'{e}, , France, 
$^5$INFN Bari, Italy, 
$^6$INFN Tor Vergata, Italy, 
$^7$University of Roma Tor Vergata, Italy, 
$^8$KTH Royal Institute of Techinology, Stockholm Sweden, 
$^9$University of Catania, Italy, 
$^{10}$INFN Catania, Italy, 
$^{11}$RIKEN, Wako, Japan, 
$^{12}$OATo - INAF Turin, Italy, 
$^{13}$Russian Space Corporation Energia, Moscow, Russia, 
$^{14}$National Centre for Nuclear Research, Lodz, Poland, 
$^{15}$UTIU Rome, Italy, 
$^{16}$Omega, Ecole Polytechnique, CNRS/IN2P3, Palaiseau, France, 
$^{17}$INFN - Laboratori Nazionali di Frascati, Italy, 
$^{18}$Konan University, Japan 
}

\forColl{JEM-EUSO} 

\emailAdd{francesco.fenu@unito.it}
\emailAdd{miyamoto@to.infn.it}

\abstract{
Mini-EUSO is a mission of the JEM-EUSO program flying onboard the International Space Station since August 2019. 
Since the first data acquisition in October 2019, more than 35 sessions have been performed for a total of 52 hours of observations. 
The detector has been observing Earth at night-time in the UV range and detected a wide variety of transient sources all of which have been modeled through Monte Carlo simulations. 
Mini-EUSO is also capable of detecting 
meteors and potentially space debris 
and we performed simulations for such events to estimate their impact on future missions for cosmic ray science from space.
We show here examples of the simulation work done in this framework to analyze the Mini-EUSO data. 
The expected response of Mini-EUSO with respect to ultra high energy cosmic ray showers has been studied. 
The efficiency curve of Mini-EUSO as a function of primary energy has been estimated and the energy threshold for Cosmic Rays has been placed to be above $10^{21}$ eV. 
We compared the morphology of several transient events detected during the mission with cosmic ray simulations and excluded that they can be due to cosmic ray showers. 
To validate the energy threshold of the detector, a system of ground based flashers is being used for end-to-end calibration purposes. 
We therefore implemented a parameterization of such flashers into the JEM-EUSO simulation framework and studied the response of the detector with respect to such sources.
}

\FullConference{37$^{\rm{th}}$ International Cosmic Ray Conference (ICRC 2021)\\
		July 12th -- 23rd, 2021\\
		Online -- Berlin, Germany}

\begin{document}
\maketitle
\vspace{-0.1cm}
\section{Introduction}
\begin{figure}
\vspace{-0.2cm}
\centering
\includegraphics[width=0.6\columnwidth]{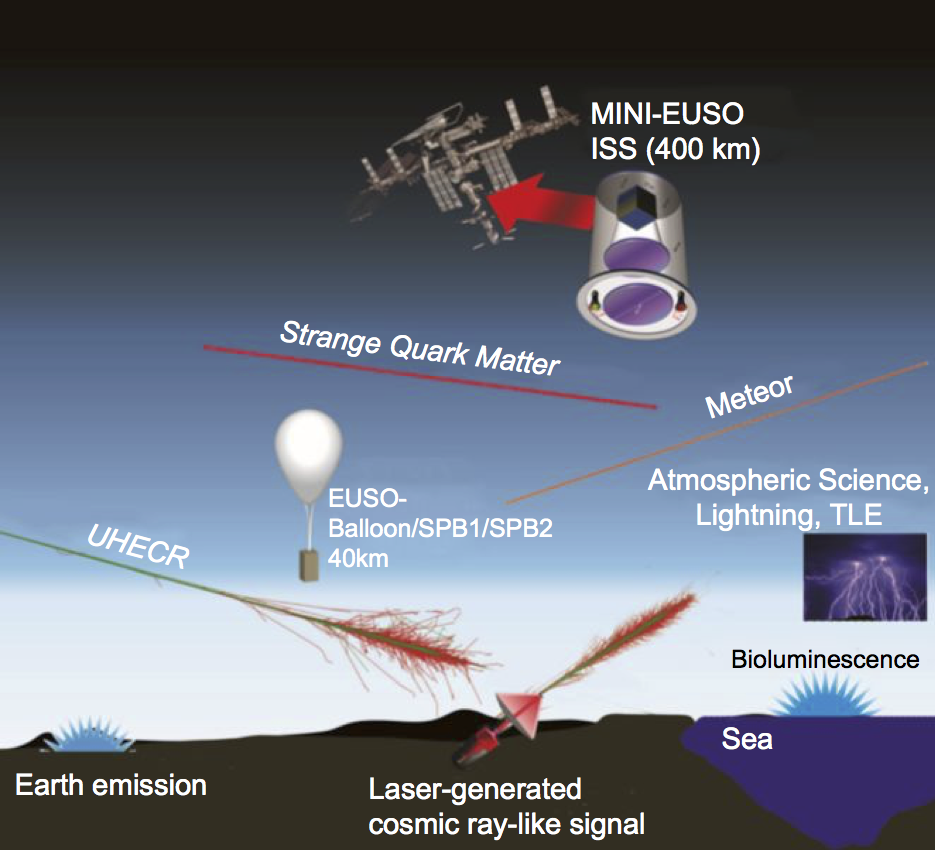}
\caption{
  Mini-EUSO mission summarised in one diagram. 
  From the ISS, Mini-EUSO will observe a variety of phenomena in the UV range, 
  in addition to creating a high resolution UV map of the Earth.
  \vspace{-0.3cm}
}
\label{fig:program}
\end{figure}
\vspace{-0.3cm}
Mini-EUSO~\cite{ref:MElaunch} is a scientific mission within the JEM-EUSO program~\cite{ref:JEM-EUSO}.
The telescope has been launched in August 2019 and currently in operation onboard the International Space Station (ISS).
The main goal of Mini-EUSO is to measure the UV emissions from the ground and atmosphere, using an orbital platform.
These observations will provide interesting data for the scientific study of a variety of UV phenomena such as transient luminous events (TLEs), meteors, space debris, hypothetical strange quark matter (SQM) and bioluminescence~\cite{ref:MEscience}, as summarised in Fig. ~\ref{fig:program}.
Moreover, this will allow us to characterise the UV emission level, which is essential for the optimisation of the design of future EUSO instruments for Extreme-Energy Cosmic Ray (EECR) detection.
Mini-EUSO observes the atmosphere from a nadir-facing window inside the Zvezda module of the ISS.
It is based on one EUSO detection unit, referred to as the Photo Detector Module (PDM).
The PDM consists of 36 Multi-Anode Photomultiplier Tubes (MAPMTs), each one having 64 pixels, for a total of 2304 pixels.
The MAPMTs are provided by Hamamatsu Photonics, model R11265-M64, and are covered with a 2 mm thickness of BG3 UV filter with anti-reflective coating.
The full Mini-EUSO telescope consists of 3 main systems: the optical system, the PDM and the data acquisition system~\cite{ref:MEelectronics}.
The optical system of 2 Fresnel lenses is used to focus light onto the PDM in order to achieve a large field of view (FoV, $44^\circ\times44^\circ$) with a relatively light and compact design, well-suited for space application~\cite{ref:MEoptics}.
The PDM detects UV photons and is read out by the data acquisition system with a sampling rate of 2.5$~\mu$s and a spatial resolution of $\sim$6 km.
\vspace{-0.2cm}
\section{Simulations of typical observations}
\begin{figure}
\centering
\includegraphics[width=0.7\columnwidth,height=5cm]{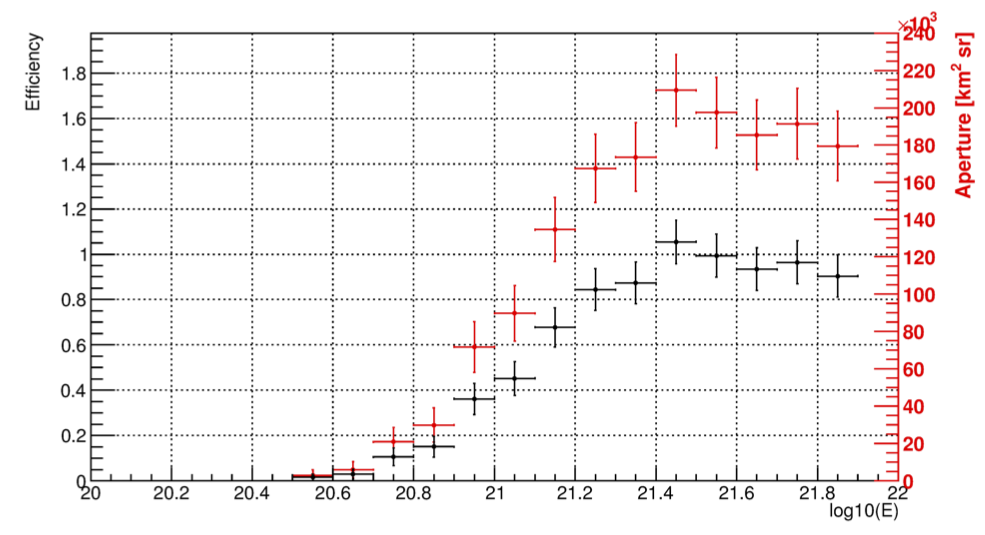}
\caption{
  The detection exposure efficiency (on the left axis, in black) and geometrical 
  aperture (on the right axis, in red) are shown as a function of the EAS energy logarithm. 
  A UV background level of 1 photon/pixel/GTU was considered in both cases.
  \vspace{-0.3cm}
}
\label{fig:exposure}
\end{figure}

\vspace{-0.2cm}
The Mini-EUSO configuration has been included in the EUSO Simulation and Analysis Software (ESAF) package~\cite{ref:MEesaf}.
ESAF is one of the official software tools to perform simulations of Extensive Air Shower (EAS) development, photon production and transport through the atmosphere and detector response for optics and electronics.
Moreover, ESAF contains algorithms for the reconstruction of the properties of EAS produced by EECRs.
Originally developed for the ESA-EUSO mission, all the planned missions of the JEM-EUSO program have been implemented in ESAF in order to assess the full range of expected performance for cosmic ray observation.
Although Mini-EUSO is not designed to detect EECRs due to the small size of the optical system, it will be possible to provide an upper limit for a null detection with 
its large geometrical aperture 2$\cdot10^5 km^2$sr above $10^{21}$ eV (see Fig.\ref{fig:exposure}).
A UV background level of 1 photon/pixel/GTU (Gate Time Unit, 1 GTU=2.5$\mu$s) was considered, which corresponds to a UV nightglow intensity of $\sim$500 photons/$m^2$/ns/sr, the typical value expected on oceans during dark nights~\cite{ref:MEtelescope}.
Additionally, during flight it will be possible to simulate EECR-like signals using ground-based laser facilities in order to verify the capability of Mini-EUSO to detect cosmic rays and to allow the testing and optimisation of the trigger system.
Also, several flashers based on powerful white or UV LEDs are developed in Italy, Japan and Russia and performed a couple of campaigns as described later.
\subsection{Natural UV light sources in the atmosphere}
\vspace{-0.2cm}
In the past years, various atmospheric phenomena such as TLEs and blue jets
as well as meteors 
have been simulated.
The former ones have typical duration of tens of ms
while the latter have longer duration ranging from the order of a second to minutes.
TLEs have UV luminosity and high frequencies~\cite{ref:TLE2}, and thus should be well characterised to avoid interference with EECR detection and triggering.
Mini-EUSO has a dedicated trigger algorithm to capture TLEs and other millisecond scale phenomena at high resolution~\cite{ref:multitrigger}.
These could help improve the understanding of the formation mechanisms of filaments plasma structures, complementing atmospheric science experiments.
The TLEs simulated in the ESAF is described here~\cite{ref:MEesaf}.\\
Mini-EUSO is also capable of detecting slower events such as meteors, fireballs, strange quark matter (SQM) and space debris with magnitudes of M $<$ +5.
In optimal dark conditions, the signal (integrated at steps of 40.96 ms) will exceed the UV-nightglow level by 3-4$\sigma$.
These events will be detected using offline trigger algorithms on ground~\cite{ref:Lech}.
The implementation of the meteor phenomena in ESAF is described in~\cite{ref:meteoresaf}, 
which inherits the approach described in~\cite{ref:meteoresaf2}.\\
\begin{figure}
        \begin{center}
          \includegraphics[width=0.9\columnwidth]{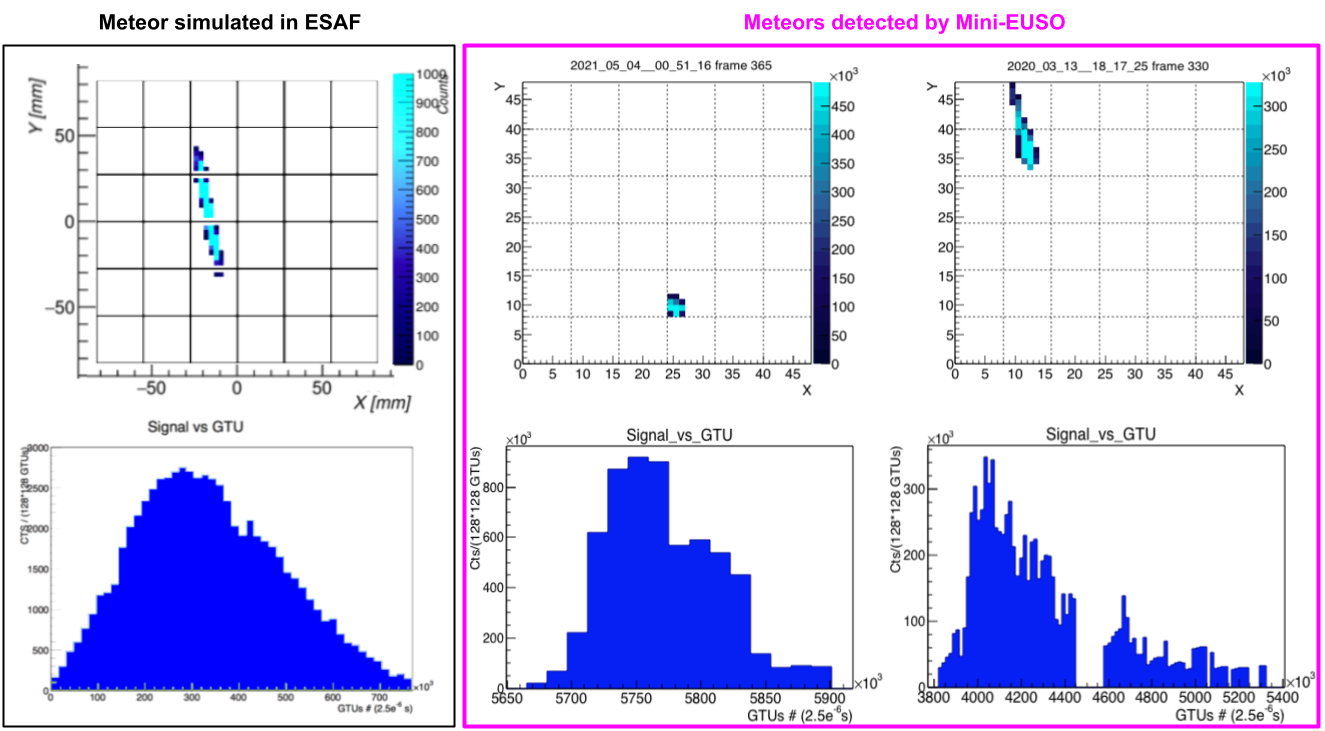}
          \caption{
            Left:Expected light track of a meteor of absolute magnitude M = +5 
            detected by Mini-EUSO (the effects of UV-nightglow are not included and a threshold has
            been applied at 30 counts). 
            Bottom: Expected light profile. 
            Each time bin on the x-axis corresponds to an integration time of 40.96 ms, the resolution of the level 3 data from Mini-EUSO.
            Centre \& right: Example of meteors detected by Mini-EUSO. 
            In the Mini-EUSO data, there are meteors with different brightness and time duration.
            Further analysis is currently ongoing.
            \vspace{-0.5cm}
          }
          \label{fig:meteor}
        \end{center}
\end{figure}
Fig.\ref{fig:meteor} shows examples of a meteor track.
Left plots shows a meteor event simulated in ESAF having absolute magnitude M = +5 crossing the FoV of Mini-EUSO with a 45$^\circ$ inclination with respect to the nadir axis. 
The meteor speed is 70 km/s and its duration is 2 s.
The right panel shows two examples of meteor events detected by Mini-EUSO.
The duration of events are $\sim$0.6 s and $\sim$3.8 s for each.
The blank part in the middle of light curve in the bottom is due to the gap between 2 PMTs where meteor is passing through.
In the Mini-EUSO data, there are meteors with different brightness and time duration and the detailed analysis is currently ongoing.
\subsection{Space Debris}
\begin{figure}[h]
        \begin{center}
           \includegraphics[width=0.9\columnwidth]{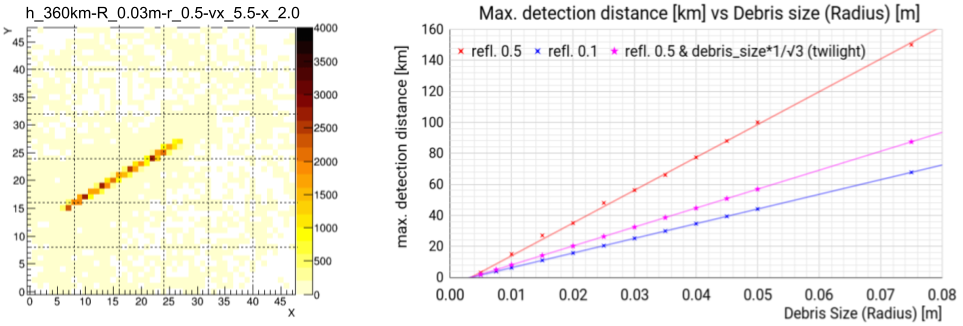}
          \caption{
            Space debris simulated in ESAF and maximum distance of space debris observable by Mini-EUSO from the ISS as a function of reflectance and size of debris derived from simulation study with ESAF.
            \vspace{-0.5cm}
          }
          \label{fig:debris}
        \end{center}
\end{figure}
A detector like Mini-EUSO is also potentially capable of detecting space debris.
If it were with a shade to prevent direct sunlight reaching the lens without hindering the FoV,
Mini-EUSO would be effectively a high-speed camera with a large FoV and can be used as a prototype for the detection of space debris during the twilight periods of observation~\cite{ref:SD2019}.
It will detect debris when they are illuminated by the Sun, but the instrument is in darkness.
In the current simulation with ESAF, the photon flux of the Sun in the 300-400 nm has been considered.
The debris are assumed to have a spherical shape of diameter d and a variable reflectance.
The Fig.\ref{fig:debris} shows the potential of Mini-EUSO to detect space debris.
The condition set in the simulation is that the signal is at least 3$\sigma$ above background for at least 5 consecutive blocks of 40.96 ms each.
The UV background has been assumed to be 
the same as in other simulations, 
which is 1 count/pixel/GTU, 
typical value expected on oceans during dark nights.
However, it is possible that for this specific measurement, the background could be higher due to the presence of some sunlight.
The detectable debris size will be changed depending on the background level at dawn on orbit which is under study.
During the pre-flight tests using Mini-EUSO Engineering Model (Mini-EUSO EM) in 2018~\cite{ref:MEEM}, we observed that the level of background photons increased by less than a factor of 3 during the twilight time towards dawn.
This indicates that the signal decreases by less than a factor 3. 
Assuming that this value is appropriate also for the space observation, 
it corresponds to the fact that detectable debris size decreases by less than a factor of $\sqrt{3}$. 
The magenta coloured plot in the Fig.\ref{fig:debris} shows the maximum detectable distance by Mini-EUSO-like telescope in the case of reflectance of 0.5 with taking into account increasing background photons in twilight time.
\subsection{Flashers}
\begin{figure}[h]
        \begin{center}
           \includegraphics[width=0.96\columnwidth]{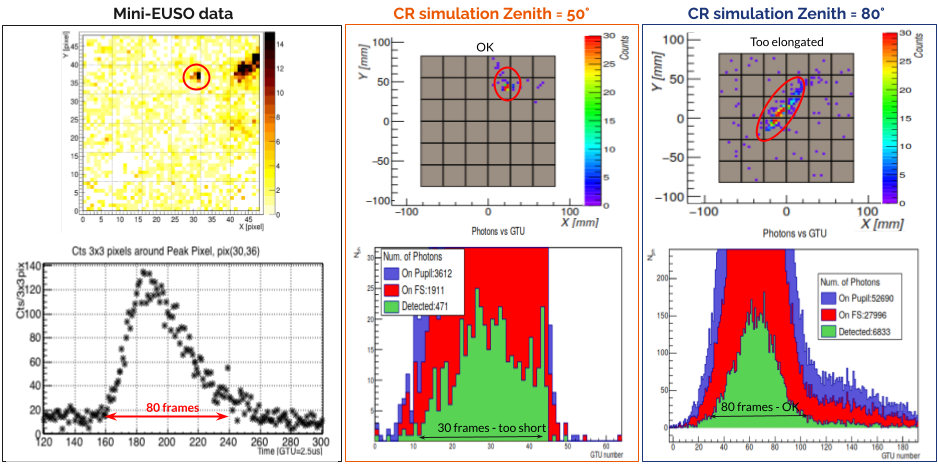}
          \caption{
            Focal plane view and light curve of Mini-EUSO signals and simulations. 
            Left: Mini-EUSO event detected off the coast of Sri Lanka. 
            Centre and Right: EAS simulated through ESAF with different energy and zenith angle. 
            The simulation with $Z=50^{\circ}$ and energy $5\times10^{21}$ eV produce a footprint on the focal plane similar to the event but the light curve is too short, while the event at $Z=80^{\circ}$ and energy $2\times10^{22}$ eV correctly reproduce the light curve but has a different shape on the focal plane.
            \vspace{-0.6cm}
          }
          \label{fig:flasherEvent}
        \end{center}
\end{figure}
Mini-EUSO has detected a flasher light on the ground off the coast of Sri Lanka (left panel of Fig.\ref{fig:flasherEvent}).
We simulated EAS events in ESAF with different energy and zenith angle~(Z) for a comparison.
As shown in the right panel of Fig.\ref{fig:flasherEvent},
the simulation with $Z=50^{\circ}$ and energy $5\times10^{21}$ eV produces a footprint on the focal plane similar to the event but the light curve is too short, while the event at $Z=80^{\circ}$ and energy $2\times10^{22}$ eV correctly reproduces the light curve but has a different shape on the focal plane.
In such a way ESAF simulation is used also to verify the detected events.
\vspace{-0.1cm}
\subsubsection*{Flasher campaign in May 2021}
\begin{figure}[h]
  \begin{center}
    \includegraphics[width=1\columnwidth]{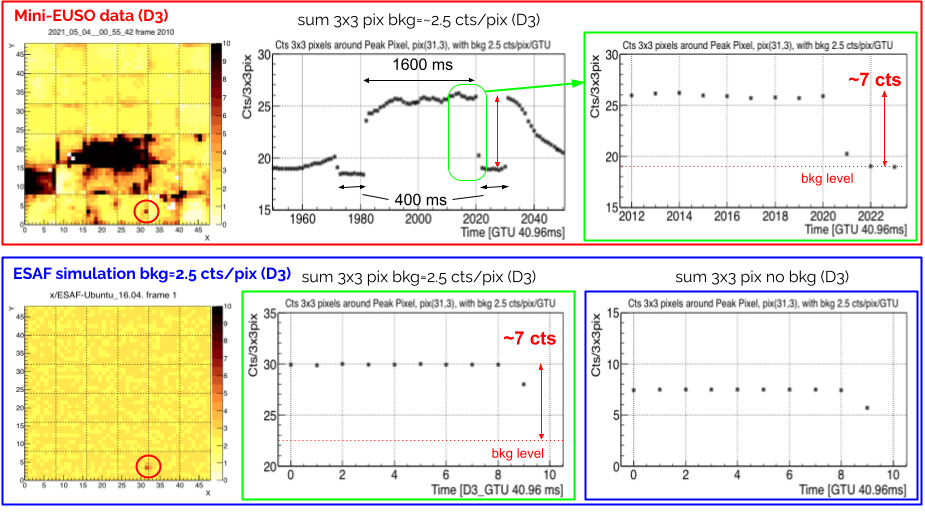}
    \caption{
      Top: Mini-EUSO flasher campaign event. 
      Center shows the light evolution of summed counts of 3 by 3 pixels around the peak count pixel, indicated by red-circle on the left plot for a cycle of 1600 ms pulse.
      Transit of the 1600 ms pulse with 400 ms off before and after the pulse is clearly seen.
      Right shows the zoomed plot to the duration and timing where ESAF simulated the same event as shown in the bottom plots as following.\\
      Bottom: Reproduced flasher campaign event by ESAF.
      Center shows the light evolution of summed counts of 3 by 3 pixels around the peak count pixel, indicated by red-circle on the left plot with a background level of 2.5 cts/pix/GTU.
      Right shows the same as the center without background to see the signal counts clearly.
      \vspace{-0.5cm}
    }
    \label{fig:UVLEDflasher}
  \end{center}
\end{figure}

\vspace{-0.1cm}
Several kinds of ground based flashers have been developed by some groups of Mini-EUSO collaboration such as in Japan, Italy and Russia.
One of such flashers was developed in Turin and tested on the ground in advance by a telescope which consists of same type of Hamamatsu MAPMTs and electronics as Mini-EUSO telescope.
The flasher consists of 100W COB-UV LEDs, DC power supply and Arduino circuit. 
Taking into account that Mini-EUSO pixel FoV is $\sim$6 km passing by at a velocity of ISS which is $\sim$7.5 km/s, it will take 800 ms to pass completely one pixel.
Thus, to be sure one pixel has constant full pulse, and also so that we can see the transit of a pulse within a pixel, 
while its passing through the Mini-EUSO FoV,
LEDs were pulsed 
6 times in 12 s with a pulse of 1600 ms on and 400 ms off each, followed by 12 pulses in 9.6 s with 400 ms on and 400 ms off each.
For the flasher test in advance, 
we set our telescope in the dark environment at -4th floor of Physics department of Turin at TurLab facility~\cite{ref:TurLab}.
The distance of the telescope and one UV LED flasher is 40.6 m.
To reduce the light as well as to obtain only parallel light, we collimated the light at 30 cm distance from the detector focal surface with a pin-hole of 0.1 mm diameter.
As a result, the total number of photons we obtained is $\sim$60 cts/LED, which corresponds to 87.7 cts/pix/GTU with pile-up correction for each LEDs.
The flasher campaign has been done at Piana di Castelluccio, Italy at the 1500 m above sea level, in the clear night sky condition of May 3rd to 4th.
Implementing the elevation of flasher 
and the data taken by the telescope on the ground described above,
taking into account the difference in number of the LEDs (the number obtained here is the counts for 1 LED while we used an array of 9 LEDs for Mini-EUSO), 
we simulated the flasher campaign events in ESAF.
The top panel of the Fig.\ref{fig:UVLEDflasher} shows the Mini-EUSO data.
The left is the raw data image of one D3\_GTU (=40.96ms) frame while the center shows 
the light curve of the integrated counts of 3 by 3 pixels around the pixel indicated by the red circle, 
while the right shows the zoomed image around green circled part of the center plot 
corresponding to the timing and duration ESAF simulated for.
The bottom panel shows the simulated flasher event by ESAF with background level of 2.5 cts/pix/GTU similar to the one on the Mini-EUSO flasher data.
Since where the flasher was located is not far from urbanised area in Italy, near Rome and its suburbs, from Mini-EUSO spatial resolution point of view ($\sim$6 km), the background is higher than the typical night sky background which is 1 count/pix/GTU as mentioned above.
The center shows the integrated counts of 3 by 3 pixels around the pixel indicated by the red circle of the left plot, 
with background photons,
while the right shows the integrated counts of the same pixels but without background photons to see how many counts in total we obtained.
One may see the tendency that the count is going down as the flasher moves in the FoV in both Mini-EUSO and ESAF data.
As a preliminary result, the expected number to be detected obtained by ESAF is well matching to the detected number by Mini-EUSO.
We still need further analysis to understand, for example, the precise PSF and how many counts are lost in the gap of two PMTs.
We will also calibrate either or both of the telescope on the ground or flasher LEDs in the future so that we will achieve an end-to-end calibration of Mini-EUSO.
\vspace{-0.2cm}
\section{Conclusions}
\vspace{-0.2cm}
The Mini-EUSO detector has been successfully implemented in ESAF.
Several kinds of atmospheric UV phenomena such as TLEs, blue jets and meteors, as well as space debris 
have been simulated by ESAF and compared to the Mini-EUSO data.
Several kinds of UHECR events are simulated by ESAF and used to verify a ground-based flasher event observed by Mini-EUSO.
The UV LED flasher campaign event has been also simulated using the photon counts data taken by a telescope on the ground, 
which consists of the same type of detector and electronics as Mini-EUSO detector. 
The next step is the absolute calibration of either or both of the detector and the flashers on the ground.
Further detailed study on ESAF by means of the advantage of having real data obtained by Mini-EUSO onboard the ISS is ongoing to improve the simulation accuracy for better estimation of detector performance, verification of natural and artificial light sources, and to achieve the end-to-end calibration with ground based flashers and lasers.
\vspace{-0.3cm}
\section{Acknowledgements}
\vspace{-0.3cm}
This work was supported by State Space Corporation ROSCOSMOS, by the Italian Space Agency through the ASI INFN agreement n. 2020-26-HH.0 and contract n. 2016-1-U.0, by the French Space agency CNES, National Science Centre in Poland grant 2017/27/B/ST9/02162.\\
This research has been supported by the Interdisciplinary Scientific and Educational School of Moscow University ``Fundamental and Applied Space Research''. The article has been prepared based on research materials carried out in the space experiment ``UV atmosphere''.

%
%
\clearpage
\section*{Full Authors List: \Coll\ Collaboration}

\begin{sloppypar}
{\small \noindent
G.~Abdellaoui$^{ah}$,
S.~Abe$^{fq}$,
J.H.~Adams Jr.$^{pd}$,
D.~Allard$^{cb}$,
G.~Alonso$^{md}$,
L.~Anchordoqui$^{pe}$,
A.~Anzalone$^{eh,ed}$,
E.~Arnone$^{ek,el}$,
K.~Asano$^{fe}$,
R.~Attallah$^{ac}$
% ...
}
\end{sloppypar}
\vspace*{.3cm}

{ \footnotesize
\noindent
% ...
$^{ac}$ LPR at Dept. of Physics, Faculty of Sciences, University Badji Mokhtar, Annaba, Algeria \\
% ...
$^{ah}$ Telecom Lab., Faculty of Technology, University Abou Bekr Belkaid, Tlemcen, Algeria \\
% ...
$^{cb}$ Universit\'e de Paris, CNRS, AstroParticule et Cosmologie, F-75013 Paris, France\\
% ...
$^{ed}$ Istituto Nazionale di Fisica Nucleare - Sezione di Catania, Italy\\
% ...
$^{eh}$ INAF - Istituto di Astrofisica Spaziale e Fisica Cosmica di Palermo, Italy\\
% ...
$^{ek}$ Istituto Nazionale di Fisica Nucleare - Sezione di Torino, Italy\\
$^{el}$ Dipartimento di Fisica, Universita' di Torino, Italy\\
% ...
$^{fe}$ Institute for Cosmic Ray Research, University of Tokyo, Kashiwa, Japan\\
% ...
$^{fq}$ Nihon University Chiyoda, Tokyo, Japan\\
% ...
$^{md}$ Universidad Polit\'ecnia de madrid (UPM), Madrid, Spain\\
% ...
$^{pd}$ University of Alabama in Huntsville, Huntsville, AL; USA\\
$^{pe}$ Lehman College, City University of New York (CUNY), NY, USA\\
% ...
}
%
%
%
%
%
%

\end{document}